# Pressure-induced Superconductivity in Crystalline Boron Nanowires

Liling Sun[1*], Takahiro Matsuoka[2], Yasuyuki Tamari[2], Katsuya Shimizu[2*], Jifa Tian[1], Yuan Tian[1], Chendong Zhang[1], Chengmin Shen[1], Wei Yi[1], Hongjun Gao[1], Jianqi Li[1]，Xiaoli Dong[1] and Zhongxian Zhao[1]

[1] Institute of Physics and Beijing National Laboratory for Condensed Matter Physics, Chinese Academy of Sciences, Beijing, 100190, P. R. China
[2] Center for Quantum Science and Technology under Extreme Conditions, Osaka University, 1-3 Machikaneyama, Toyonaka, Osaka, 560-8531, Japan

We report high-pressure induced superconductivity in boron nanowires (BNWs) with rhombohedral crystal structure. Obviously different from bulk rhombohedral boron, these BNWs show a semiconductor-metal transition at much lower pressure than bulk boron. Also, we found these BNWs become superconductors with Tc=1.5 K at 84 GPa, at the pressure of which bulk boron is still a semiconductor, via in-situ resistance measurements in a diamond anvil cell. With increasing pressure, Tc of the BNWs increases. The occurrence of superconductivity in the BNWs at a pressure as low as 84 GPa probably arises from the size effect.

It is well-known that bulk solid boron is a semiconductor with a band gap of ~2 eV at ambient pressure. Theoretical calculations predicted that at sufficient high pressure, band overlap occurs, which drives the bulk boron to a poor metal [1-2]. Recently, two striking resistance measurements under high pressure found that the semiconductor-metal transition occurs at room temperature in bulk β-boron at 130 GPa [3] and in bulk α-boron at 160 GPa [4]. At low temperature, they both showed a superconducting transition at ~4 K and 160 GPa. Nanomaterials with the same crystal structure as the corresponding bulk solid are expected to have interesting physical properties in comparison with their bulk counterparts. For example, when the size of the solid is small enough, solid-solid phase transition pressures vary with size change [5-6]. Superconducting properties are also altered when the effective size of a superconductor is reduced [7-16]. Progress in preparation of boron nanomaterials [18-19] motivated this investigation of solid-solid phase transition and physical properties at ambient and at high pressures. In this study, we report observations of semiconductor-metal-superconductor transitions in crystalline boron nanowires (BNWs) under high pressure. In addition, the pressure dependence of the superconducting critical temperature (Tc) was studied up to 240 GPa. The correlation between the measured value of Tc and pressure in BNWs is compared with the data of bulk solid boron.

Bulk solid boron has a variety of phases, including α- rhombohedral $B_{12}$ (α-$B_{12}$), α-tetragonal $B_{50}$ (α-$B_{50}$) and β- rhombohedral $B_{105}$ (β-r-B) [17,20]. All forms mentioned above have the common structural component of boron icosahedrons $B_{12}$ in the unit cell [21]. To determine the structural properties, the BNW samples fabricated by chemical vapor deposition [19] were characterized using scanning electron microscopy (SEM), transmission electron microscopy (TEM), electron energy loss spectroscopy (EELS) and X-ray diffraction (XRD) measurement. Figure 1(a) shows the SEM image of BNWs stripped off the Si substrate. The diameter of the wires varies from 200 to 500 nm and the length from 20 to 50 μm. High resolution TEM observations on the nano-samples reveal the presence of complex structural features, as typically illustrated in Figure 1(b). From a partial enlargement of Figure 1 (b), the notable contrast anomalies in association with structural distortion, indicated by arrows, exclude the possibility of stacking faults, as shown in Figure 1 (c). Actually alterations of crystallographic orientation can be clearly recognized crossing certain defective regions, as displayed in Figure 1 (b). These facts suggest that each nanowire sample observed from SEM is texturally made up from many slim nanowires with diameter ~30 nm. To determine the composition, we analyzed the slim nanowires using EELS. Figure 1(d) exhibits the representative EELS of a slim wire sample. The boron absorption features at the K shell ionization edge (~190 eV) are clearly seen, which indicates that no other elements or impurities are observed in the BNWs. XRD measurements demonstrate that all peaks of the BNWs can be indexed to β-rhombohedral ($B_{105}$) structure, as shown in Figure 1(e). This indicates that the BNWs used in this study have the same crystal structure as bulk β-r-B [3].

High-pressure experiments were performed using a diamond anvil cell made of Be-Cu alloy. Diamonds were selected carefully for very low birefringence with tips whose diameter is 300 μm for the first experiment and 40 μm for the second. A standard four-lead technique was used for the first experiment below 60 GPa and a pseudo four-lead technique was used for the second experiment. The pseudo four-lead pattern was made by thin-film fabrication and photolithography technology. A layer of titanium (Ti) was deposited onto the diamond surface, followed by a layer of platinum deposited over the Ti layer. Four layered leads were then connected to 5μm-thick Pt wires. Insulation from the rhenium gasket was achieved by a thin layer of a mixture of diamond powder and epoxy. The size of each wire sample used in this study was measured under a high magnification observation microscope. Seven well-aligned wire samples were placed on the top anvil and then pressed into the insulating gasket with leads. No pressure medium was used. The pressures were determined at room temperature by ruby fluorescence method [22-23] and at low temperature with the diamond Raman shift [24-25]. Superconducting transitions under pressure were measured in a $^3$He/$^4$He Dilution Refrigerator.

Figure 2 shows the pressure dependence of resistivity ($\rho$) of the BNWs for two individual experiments, which were carried out at room temperature. Unlike bulk β-r-B [3], the resistance of the sample is measurable (~83 KΩ) at 300 K and 0.7 GPa. With increasing pressure, the $\rho$ value decreased significantly at 28 GPa and is saturated at this value at higher pressures. The $\rho$ value of the BNWs at 28 GPa is about $2\times10^{-3}$ Ω cm (corresponding conductivity $\sigma=500$ $\Omega^{-1}cm^{-1}$) which is close to that of minimum metallic conductivity [26,3]. The results suggest that the BNWs become a poor metal at this pressure. At 84 GPa the resistance plunges at 1.5 K, as shown in Figure. 3. The abrupt drop in resistance from a finite value at 1.5 K and a pressure of 84 GPa is a sign of superconducting transition. To confirm the resistance drop at 1.5 K is related to the superconducting transition of the BNWs, the resistance versus temperature is measured at different magnetic fields and at the fixed pressure. The resistance curve of the sample is magnetic field dependent. The resistance drop is suppressed by an applied magnetic filed and disappears at 2.5 T. It is known that the observed resistance (Ro) is composed of three parts, sample resistance (Rs), contact resistance between the sample and leads (Rc) and deformation resistance (Rd), i.e. Ro=Rs+Rr (Here Rr is residual resistance, Rr=Rc+Rd). As the resistance of the sample was measured with pseudo four-lead (two-point contact) technique, T-independent behavior with Ro ~ 50 Ohm at T < 0.5 K (as shown in Figure 3) indicates zero resistance of the sample. According to our previous experiments, no superconducting was observed from the same leads at temperature down to 20 mK in megabar pressure range. Therefore, the significant resistance drop at 1.5 K is unambiguously assigned to a superconducting transition of the sample.

Figure 4 (a) shows the resistance (R) of the BNWs versus temperature (T) at selected pressures. The shift of the R-T curve toward high temperature gives evidence that the critical temperature of superconducting transition (Tc) of the BNWs is enhanced with increasing pressure. Meanwhile, we note that the resistance of BNWs in their normal state increases with increasing pressure, the reason for which is that the value of Rd increase is bigger than that of Rc decrease (generally Rc decreases with pressure), as a result, the observed resistance Ro elevates. The pressure dependence of Tc is plotted in Figure 4 (b). Here Tc is determined by the onset transition temperature. For comparison, the results from Eremets et al [3] are plotted as open squares in Figure 4(b). Interestingly, the onset pressure (84 GPa) for the superconducting transition in the BNWs is much lower than that of the bulk boron (160 GPa). The Tc of BNWs increases with increasing pressure. Above 160 GPa, Tc of the BNWs does not increase up to the data of bulk β-r-B. Rather, it has a linear behavior up to 200 GPa. Fitting the data of the BNWs from 84 to 200 GPa gives a pressure coefficient dTc/dP=0.02 K/GPa. This value is lower than that (0.16 K/GPa) of bulk β-r-B. However, the dTc/dP of the BNWs was enhanced as pressure further increases. Fitting to data measured from 200 to 240 GPa gives dTc/dP=0.04 K/GPa which approaches the value (0.05 K/GPa) of bulk β-r-B achieved at pressure of 178-250 GPa [3].

The apparent discrepancy between the BNWs and the bulk β-r-B in the pressures of the semiconductor-metal-superconductor transitions may probably be attributed to size effect, because the BNWs and bulk β-r-B are in the same crystal structure, the difference between them is only size. To prove that the metallization and superconducting transition occurred at lower pressure is related to the size effect, parallel experiments on transport properties of the BNW have been carried out at ambient pressure. Figure 5 shows experimental measurements of conductivity (σ) of the BNW as a function of temperature (T). According to early studies [27-28], the boron-rich materials have a Mott's variable-range hopping (VRH) conduction nature [29-30]: $\sigma = \sigma_o \exp[-(T_o/T)^Q]$ ( $T_o = \dfrac{60}{\pi \xi^3 k_B N(E_F)}$ ) where σ is the conductivity, $T$ is the temperature, $\xi$ is the localization length of the wave function of carriers, $k_B$ is the Boltzmann constant, $N(E_F)$ is the density of state at the Fermi level and $\sigma_o$ is a constant. On the basis of Mott's theory we can estimate the $N(E_F)$ from the measured σ. It is noted for that the estimation of $N(E_F)$ is completely based on parameters of $Q$ and $T_0$ of Mott's theory, if the localization length remains unchanged. $T_0$ is determined both by $Q$ value and σ value, the latter is experimental result. Therefore, $Q$ value should be the key factor for the estimation of $N(E_F)$. In Mott's equation, $Q$ value is alterable, $Q$=1/4 for the three dimensional sample and $Q$=1/3 for the two dimensional sample. To make the BNW and bulk β-r-B comparable, we take value of $Q$ and $\xi$ for the BNW same as that ($Q$=1/4, $\xi$=0.1 nm) of bulk β-r-B in the calculations. Using this model, we obtained $T_o$=2.5×10$^7$ K and then estimated $N(E_F)$ of the BNW at ambient pressure to be 8.9 /eV nm$^3$. In order to make a precise comparison, we looked up the conductivity data for single crystal of bulk β-r-B (the BNW is a single crystal) from Ref. 31, and we found the $N(E_F)$ of the single crystal of bulk β-r-B is about 2.5 /eV nm$^3$. Comparing with the $N(E_F)$ of single crystal of β-r-B and the BNW, the $N(E_F)$ of the BNW is still higher than that of bulk β-r-B. The same phenomenon also has been found in boron nanobelt [28]. Therefore, the size effect could be the reasons for that the BNW is easier to be metallized than the bulk β-r-B under high pressure at room temperature.

We compared effect of different $Q$ values on the $N(E_F)$. Table 1 shows the model parameter $T_0$ of Mott's VRH model and $N(E_F)$ estimated with different $Q$ values. It is seen that the $N(E_F)$ increases with increasing $Q$ value. In this study, the $Q$ value of the BNW should be higher than that of bulk β-r-B, therefore, its corresponding $N(E_F)$ is also high. When the BNW is superconducting, its Tc increases gradually with increasing pressure to 160 GPa. With further increase in pressure, Tc of the BNWs continues to increase linearly up to 200 GPa, as seen in Figure 4(b). In the range of 200 to 240 GPa, the Tc and the value of dTc/dP of the BNWs approaches that of the bulk β-r-B. This means that the size effect is negligible.

In conclusion, superconductivity in BNWs with rhombohedral structure was studied under high pressure up to 240 GPa. Resistance measurements in a diamond anvil cell show that the BNWs exhibit metallization at 28 GPa at room temperature and superconductivity at 1.5 K at 84 GPa where the bulk β-r-B is still a semiconductor. It was found that pressure has a positive effect on the Tc. The pressure coefficient of the BNWs is 0.02 K/GPa over the pressure range of 84-200 GPa, followed by increasing to the value of 0.04 K/GPa from 200 GPa to 240 GPa as that of bulk β-r-B. We proposed that the size effect influences the pressure for metallization and superconducting transition of the BNWs in comparison with bulk β-r-B.

Acknowledgement

We sincerely thank Prof. W. J. Nellis of Harvard University for valuable discussions. This work was supported by Center of Excellence (COE) Project of Science Technology Agency of Japan and the National Science Foundation of China (Grant No. 50571111 and 10874230). This work was also supported by the Ministry of Science and Technology of China (2005CB724400) and the project CoMePhS.

*Corresponding author:

llsun@aphy.iphy.ac.cn
shimizu@cqst.osaka-u.ac.jp

**Figure captions**

Fig. 1 (a) SEM image of the BNWs stripped off the silicon substrate, (b) TEM image of one of wire samples used in this study, (c) Partial enlargement of Fig.1 (b), (d) EELS of the BNWs at K shell ionization edge (~190 eV), (e) X-ray diffraction spectrum of the BNWs.

Fig. 2 Resistivity (ρ) of the BNWs as a function of pressure at room temperature. The inset of the main figure shows a photograph of pseudo four-lead and seven well-aligned BNWs on the diamond tip. The diameter of the tip is 40 μm and the separation between two leads is ~4 μm.

Fig. 3 Electrical resistance (R) versus temperature (T) of superconducting BNWs at 84 GPa measured at different magnetic fields in the low T range. The inset is the R-T curve in the temperature range of 0.08-300 K under zero magnetic field.

Fig. 4 (a) Resistance-temperature curves of the BNWs at selected pressure, (b) Pressure dependence of Tc of BNWs obtained from resistance measurements. Value of Tc is determined from transition onset. Solid square represents data of this study and open square represents data from Ref. 3.

Fig. 5 Temperature dependence of conductivity shown as Ln (σ) versus (a) $T^{1/4}$ and (b) $T^{1/3}$ diagram.

Table 1 Parameter $T_0$ and $N(E_F)$ value of the BNWs with different $Q$ values. In the estimations, we take $\xi=0.1$nm.

| $T_0$ (K) | $N(E_F)$ (/eV nm$^3$) | $Q$ |
|---|---|---|
| $8.3\times10^5$ | 267 | 1/3 |
| $2.5\times10^7$ | 8.9 | 1/4 |

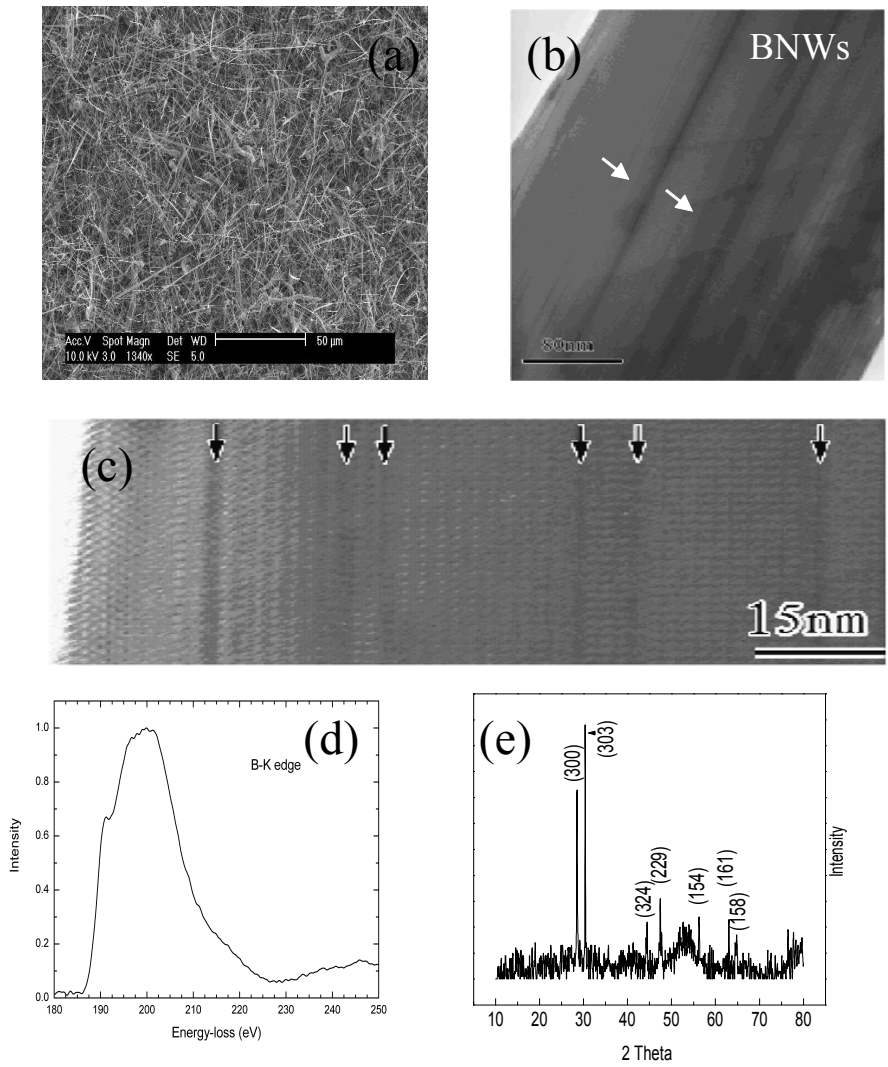

Fig.1 Sun et al

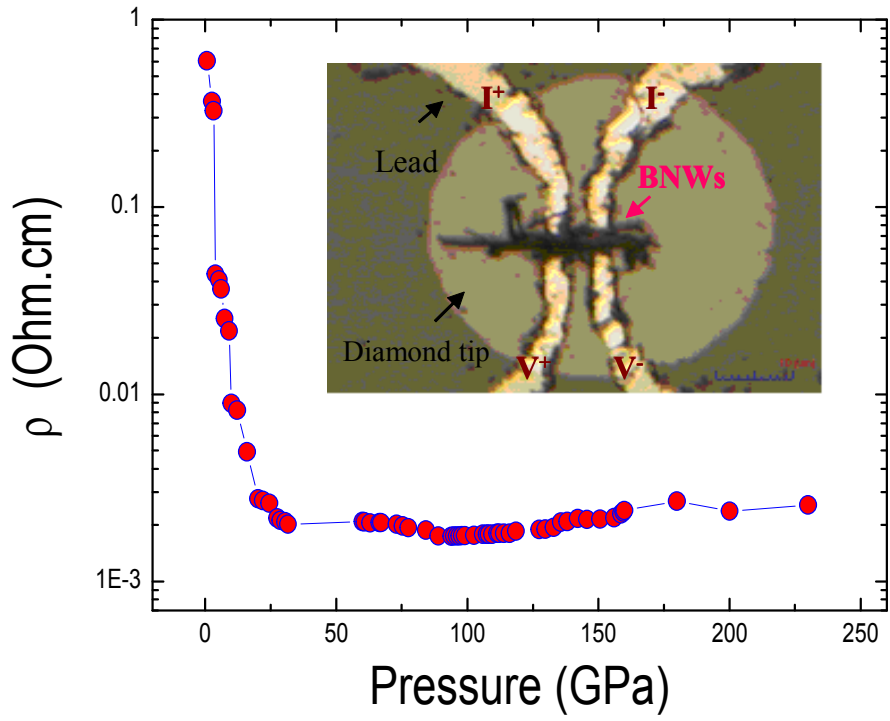

Fig. 2 Sun et al

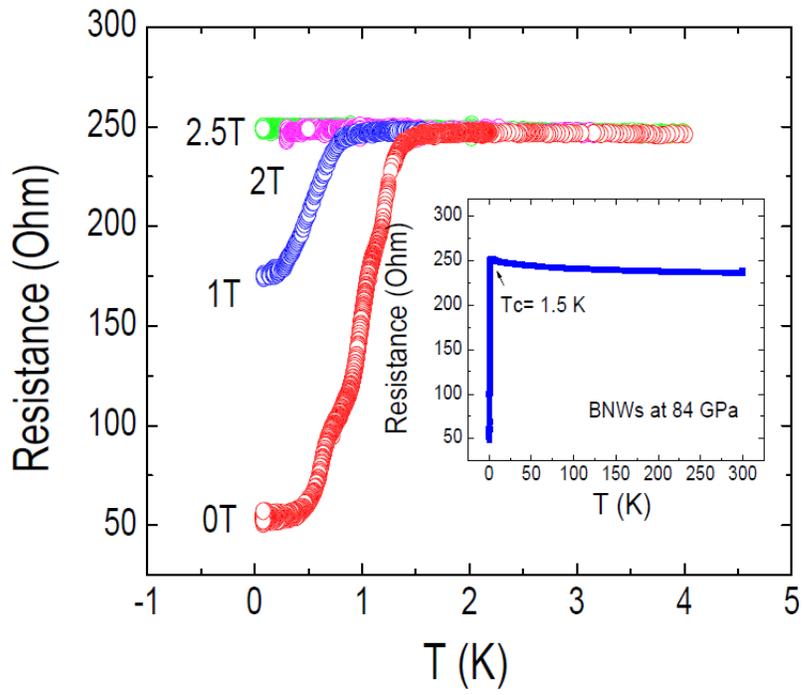

Fig. 3 Sun et al

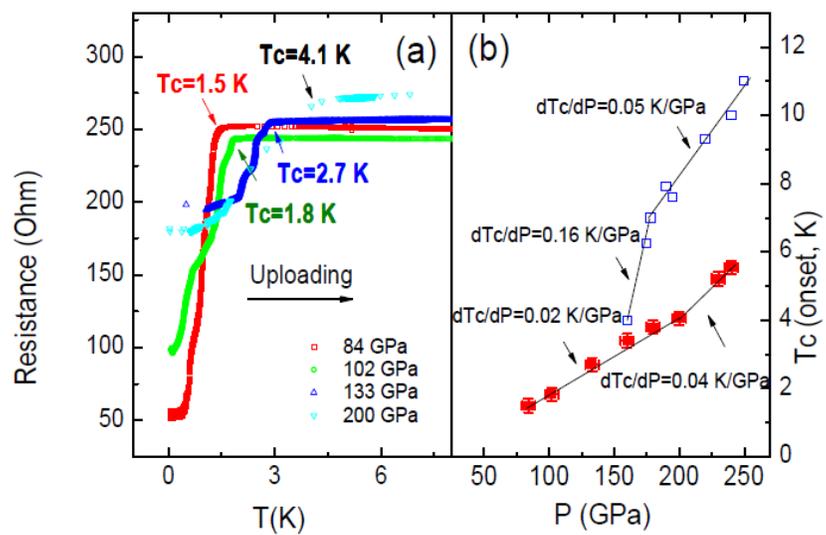

Fig. 4 Sun et al

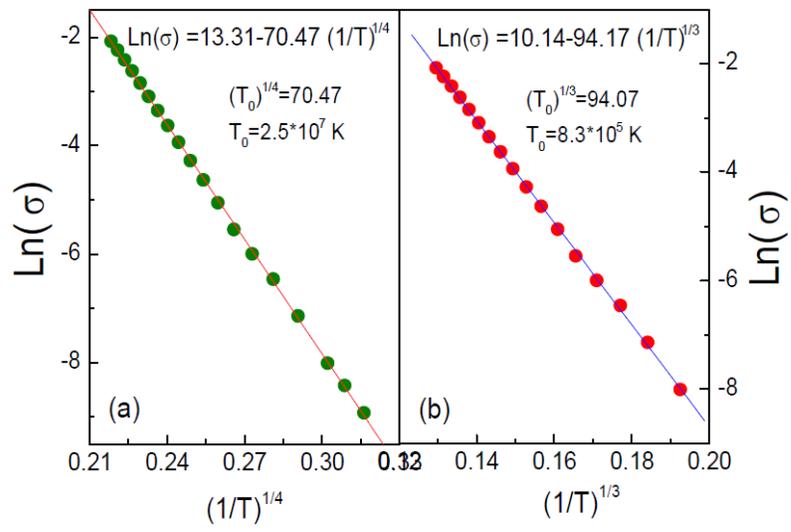

Fig.5 Sun et al